Atomic Layer Deposited Protective Coating of Aluminum Oxide on Silver-based Telescope Mirror: A Comparison Between a Pure Ozone and $H_2O$ Precursor

*Søren A. Tornøe[1], Brandon Cheney[2], Brian Dupraw[3], Yoshimasa Okamura[4], Andrew C. Phillips[3], Takayuki Hagiwara, Tetsuya Nishiguchi, Nobuhiko P. Kobayashi[1†]*

1) Søren A. Tornøe and Nobuhiko P. Kobayashi
Baskin School of Engineering, Univ. of California Santa Cruz, Santa Cruz, CA 95064, United States and Nanostructured Energy Conversion Technology and Research (NECTAR), Univ. of California Santa Cruz, Santa Cruz, CA 95064, United States

2) Brandon Cheney
Department of Earth & Planetary Sciences, University of California Santa Cruz, Santa Cruz, CA 95064

3) Brian Dupraw and Andrew C. Phillips
University of California Observatories, Univ. of California, Santa Cruz, CA, USA 95064

4) Yoshimasa Okamura
Meiden America, Inc., Santa Clara, CA 95051, United States

5) Takayuki Hagiwara and Tetsuya Nishiguchi
Meiden Nanoprocess Innovations, Inc., Tokyo, Japan

[†]Corresponding author: E-mail: nkobayas@ucsc.edu

**Keywords:**

silver, telescope mirror, e-beam evaporation, atomic layer deposition, pure ozone, thin film, high temperature, high humidity

**Abstract**

Although silver-based telescope mirrors excel over other materials such as gold and aluminum in the visible-infrared spectral range, they require robust protective coatings to overcome their inherent low durability. Our research shows that a single-layer of aluminum oxide ($AlO_x$) deposited through thermal atomic layer deposition (ALD) using trimethylaluminum (TMA) and water ($H_2O$) at low temperatures (~60°C) serves as an acceptable protective coating without adversely impacting the optical performance of the mirrors. While silver-based mirrors protected with a single-layer of $AlO_x$ perform decently in the field, in environmental tests under high-humidity at high-temperature conditions that accelerate underlying failure mechanisms, they degrade quickly, suggesting that there is room for improvement. This paper describes a study that compares the performance and endurance of two sets of silver-based mirrors protected by a single-layer of $AlO_x$ prepared by thermal ALD with two types of oxygen precursors: $H_2O$ and pure ozone (PO). The study shows that while the two types of samples, regardless of their oxygen precursors, initially have comparable spectral reflectance, the reflectance of the samples with $AlO_x$ protective coatings prepared with PO remain nearly constant 1.6 times longer than those with $AlO_x$ protective coatings prepared with $H_2O$ in the environmental test, suggesting promising characteristics of $AlO_x$ protective coatings prepared with PO.

## 1. Introduction

Silver-based telescope mirrors (Ag-based mirrors) offer optical performance in the visible to infrared spectral range that excels over that of aluminum-based mirrors, though they suffer significantly from low durability in various environment tests. Our previous study has shown that a single-layer of aluminum oxide ($AlO_x$), with approximate thickness of 60 nm, deposited through thermal atomic layer deposition (ALD) acts as a protective coating ($AlO_x$ protective coating) and significantly increases the working lifetime of Ag-based mirrors without negatively impacting the optical properties of the mirrors [1][2][3][4][5][6][7]. In designing an ALD process for any class of metal oxide, various deposition conditions such as deposition temperature, precursor pulse time, and chamber purge time are often vital for a given choice of precursors. For thermal ALD of $AlO_x$, $H_2O$ is the standard oxygen precursor that is relatively safe and easy to work with, however high deposition temperatures (100-150°C) are often required to maintain reasonable growth rates and minimize the incorporation of various contaminants [8]. In addition, $H_2O$ has a strong tendency to leave residue in the chamber, resulting in the need for longer chamber purge times after a $H_2O$ pulse. Nevertheless, the impact of choosing a particular oxygen precursor is expected to be substantial.

This paper describes a comparative study of two types of Ag-based mirrors: one with an $AlO_x$ protective coating deposited with a conventional oxygen precursor – water ($H_2O$) – and the other deposited with high purity ozone (PO) [9]. Being high purity, PO used in this study is above 80% [9], in contrast conventional ozone typically contains an $O_3$ purity below 20% [10]. PO can be used at temperatures ranging from 30-60°C which is much lower than those required for $H_2O$, as it leaves no residue in the chamber, and consequently, chamber purge times can be significantly shortened. The goal of this study is to see how well environmentally weathered $AlO_x$ protective coatings hold up when prepared with PO in comparison to those prepared with $H_2O$.

## 2. Experiment/Methods

## 2.1 Sample Preparation

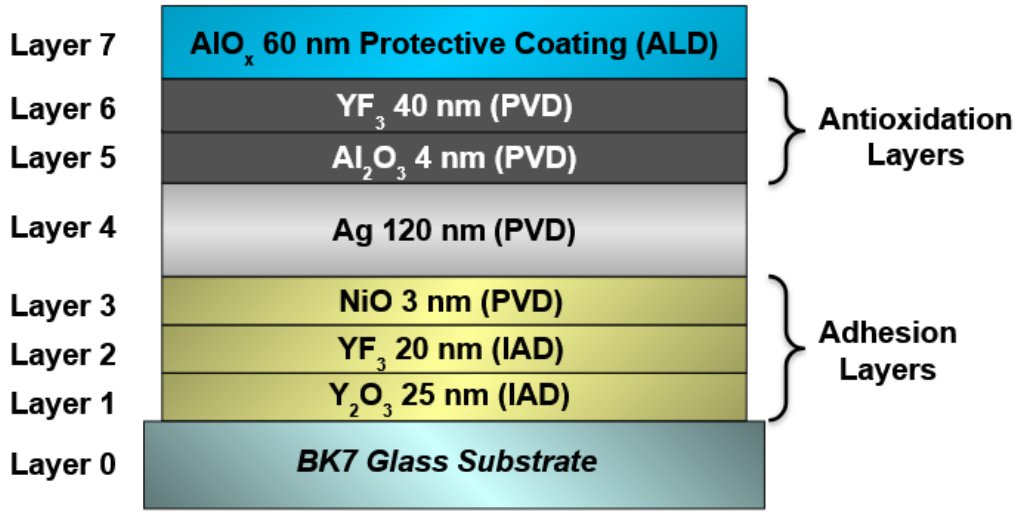

**Figure 1:** *General diagram representation of the samples produced for this study. With the exception of the protective coating, all layers were deposited via e-beam PVD.*

The Ag-based mirror samples consist of 7 layers deposited on a BK7 glass substrate as illustrated in Figure 1. Layers 1 and 2 were deposited using ion-assisted electron-beam deposition (IAD) while Layers 3 through 6 were prepared using e-beam (i.e., physical vapor deposition PVD) deposition. The IAD deposition and the e-beam deposition were carried out in a single PVD vacuum deposition system, and Layers 1 through 6 were deposited with a substrate temperature of 50℃. The adhesion layers, Layers 1 through 3, appropriately prepare the surface of the BK7 glass substrate prior to the deposition of layer 4 to ensure that the Ag sticks well to the substrate. The Ag layer has a thickness of 120 nm, which was determined to be thick enough for it to optically behave similarly to bulk polished Ag. On top of the Ag layer, two antioxidation layers, Layers 5 and 6, were deposited to temporarily protect the surface of the Ag layer during the transfer of the samples from the PVD system to an ALD system. Layer 7 was deposited via thermal ALD at 60℃ using a trimethylaluminum (TMA) precursor combined with either a $H_2O$ or PO oxygen precursor, yielding two-types of samples. Further details on all the deposition processes are presented in Table 1.

**Table 1:** *Information on individual depositions of layers for the mirror samples. The two processes – ALD with $H_2O$ and PO – were independently calibrated and optimized, such that the only difference of relevance between the two-types of samples is a choice of oxygen precursor.*

| | Material | Nominal Thickness [nm] | Deposition Rate [nm s⁻¹] | Base Pressure [Torr] | Substrate Temperature [°C] | Notes |
|---|---|---|---|---|---|---|
| Layer 7 H₂O | AlO$_X$ | 60 | 0.086 | 0.47 | 60 | Thermal ALD with TMA and H₂O precursors, N₂ carrier gas flow: 20 sccm. |
| Layer 7 PO | AlO$_X$ | 60 | 0.160 | 0.15 | 60 | Thermal ALD with TMA and PO precursors, Ar carrier gas flow: 194 sccm. |
| Layer 6 | YF₃ | 40 | 0.60 | 8.4E-7 | 50 | PVD. |
| Layer 5 | Al₂O₃ | 4 | 0.20 | 4.2E-5 | 50 | PVD with 25 sccm flow of O₂. |
| Layer 4 | Ag | 120 | 4.00 | 1.2E-6 | 50 | PVD with Ag purity of 99.99%. |
| Layer 3 | NiO | 3 | 0.20 | 9.0E-5 | 50 | PVD with 25 sccm flow of O₂. |
| Layer 2 | YF₃ | 20 | 0.36 | 1.0E-4 | 50 | IAD with 20 sccm of flow Ar. |
| Layer 1 | Y₂O₃ | 25 | 0.36 | 1.0E-4 | 50 | IAD with 36 sccm flow of O₂ and 16 sccm flow of Ar. |

## 2.2 High Temperature High Humidity Testing

The two types of mirror samples – one with an AlO$_x$ protective coating prepared with H₂O (H₂O samples) and the other with that prepared with PO (PO samples) – underwent what we have dubbed environmental testing or high temperature high humidity (HTHH) testing with the goal of comparing the two in terms of their endurance. The prepared samples (Note: each sample is 7.6 cm by 2.6 cm coupon with a thickness of 1 mm) were placed in a circular sample holder in a glass bell jar. Two dummy coupons were inserted at the beginning and end of the sample lineup to ensure that all the samples experienced identical testing conditions (i.e., the dummy coupons allowed the first and last samples in the lineup to be affected as if they were in the middle of the line up). At the start of a HTHH test cycle, 62 grams of potassium chloride and 3.1 grams of (5%) non-iodized sodium chloride were placed at the bottom of the jar. The jar was then sealed and placed in an oven set to 80°C and allowed to sit until it had reached equilibrium with the oven. Subsequently, 45 milliliters of distilled deionized water was injected onto the salt mixture via a PTFE tube inserted into the jar. The water and salt mixture ensure that the jar maintains 80% humidity for the duration of the HTHH test cycle. The tube was snapped after the injection of the water, sealing the jar for the remainder of the cycle. The samples were left under these conditions (i.e., temperature 80°C and humidity 80%) for 19 hours (i.e., a single HTHH test cycle corresponds to 19 hours), after which the samples were removed from the jar. Subsequently, spectral reflectance was taken using a Cary 5000 UV-Vis-NIR spectrophotometer across the range of 200-3500 nm. The samples were photographed with a Samsung SM-N986U – 9MP resolution – using a black velvet photography backdrop above the samples to reduce interference with photographing the sample's surface. The samples were then placed back into the jar, to repeat the test cycle.

After the samples underwent a total of 10 cycles, spectroscopic ellipsometry was performed at an incident angle of 70° on the samples using the FilmTek 4000 (Scientific Computing International, Carlsbad, California) spectroscopic reflectometry/ellipsometry system to collect the complex ratio of p polarized to s polarized reflected light – $r_p/r_s$ – via an amplitude ratio – $\mathrm{Tan}(\Psi)$ – and change in phase – $\Delta$ – of the samples between cycle 0 and cycle 10 [11]. In addition, using Q-Scope 250 Atomic Force Microscopy (AFM) with a Q-WM190 48 N/m tip, the as-prepared samples (i.e., samples that did not undergo the HTHH testing) and the samples after 10 cycles were imaged, with the resulting images interpreted using the Gwyddion data analysis software [12]. Moreover, energy dispersive spectroscopy (EDS) data was collected with acceleration voltage at 5, 7, 10, and 15 keV. All measurements, with the exception of the whole-sample photographs, were done on the specular areas (i.e., areas not visibly corroded) of the samples, as we intended to study how the original mirror surface evolved before visible corrosion emerged. With the exception of the AFM data, all the data was processed and graphed using GNU Octave [13].

## 3. Results and Discussion

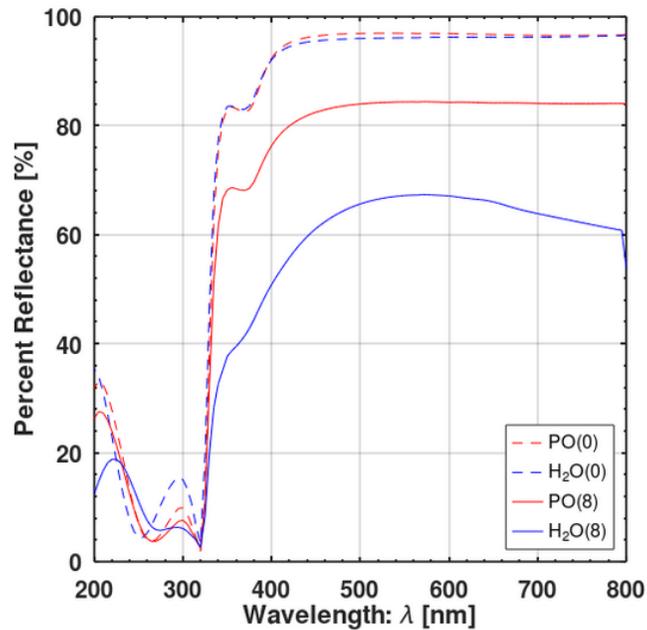

***Figure 2:*** *Reflectance spectra of PO and $H_2O$ samples: $H_2O(0)$ and PO(0) and $H_2O(8)$ and PO(8). There is little to no variation between $H_2O(0)$ and PO(0) and the variation present can easily be explained by minor inconsistencies with the evaporation process used to make the samples. After 8 cycles $H_2O$ degraded down to under 70% reflectance at 550 nm while PO only fell to 84.3% reflectance.*

Figure 2 shows the spectral reflectance of the $H_2O$ and PO samples as-prepared – $H_2O(0)$ and PO(0) – and after 8 HTHH test cycles – $H_2O(8)$ and PO(8). The comparison was made with samples after 8 HTHH test cycles instead of after 10 HTHH test cycles as they more appropriately convey how the PO samples held up in comparison to the $H_2O$ samples. The spectra of $H_2O(0)$ and PO(0) are nearly identical with minor variations (i.e., the $H_2O(0)$ spectrum is slightly lower than that of the PO(0) spectrum in the range from 400 nm to 700 nm), indicating that the deposition processes for the body of the samples (i.e., layers 1-6 in Figure 1) are highly reproducible. As illustrated in Figure 2, $H_2O(0)$ and PO(0) both had percent reflectance of approximately 97% at 550 nm. Alternatively, it could be stated that PO(0) was indeed slightly more reflective than $H_2O(0)$ with around 96% in the spectral range from 450 nm to 550 nm. After 8 HTHH test cycles, reflectance of $H_2O(8)$ significantly decreased to around 67% at 550 nm, while that of PO(8) showed considerable resilience by only degrading to around 84.3% at 550 nm and over the spectral range that was investigated. Additionally, PO(8) appears to degrade uniformly across the entire spectrum – retaining a consistent shape comparable to that of PO(0) – while $H_2O(8)$ degrades significantly more in both the ultraviolet and near infrared ranges.

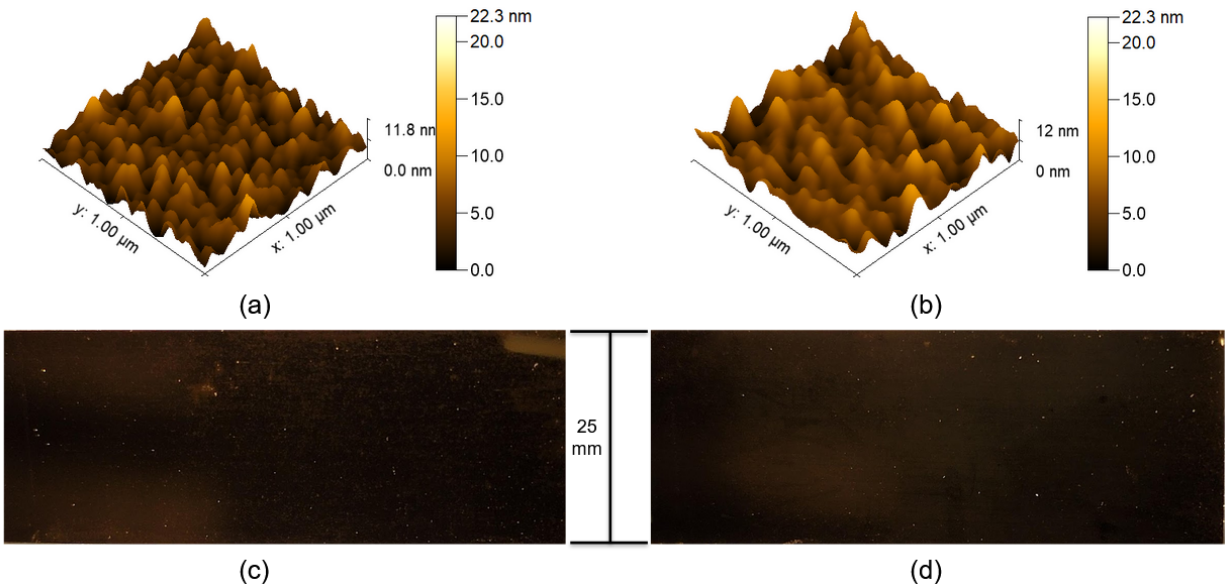

***Figure 3:*** *(a) Surface morphology of $H_2O(0)$ with RMS roughness of 1.374 nm. (b) Surface morphology of PO(0) with RMS roughness of 1.491 nm. $H_2O(0)$ has a larger number of peaks with less variation in height while PO(0) has a larger variation in peak height. (c) and (d)*

*Photos of the entire area of H₂O(0) and PO(0), respectively, where specular areas appear to have dark contrast (i.e., these as-prepared samples show an entirely specular surface).*

Figure 3 shows surface morphology collected across a specular area by AFM and optical images of the whole sample area of $H_2O(0)$ and PO(0). $H_2O(0)$ in Figure 3(a) shows general features characterized by a denser concentration of peaks and valleys while PO(0) in Figure 3 (b) shows lower density but a greater disparity between the peak heights, indicating that PO(0) is slightly rougher than $H_2O(0)$ as indicated by the RMS roughness; 1.491 nm and 1.374 nm for PO(0) and $H_2O(0)$, respectively. The respective optical images in Figure 3(c) and (d) indicate both samples are similarly reflective (i.e., specular) and show no obvious differences, indicating the two-types of samples were comparable when they were as-prepared.

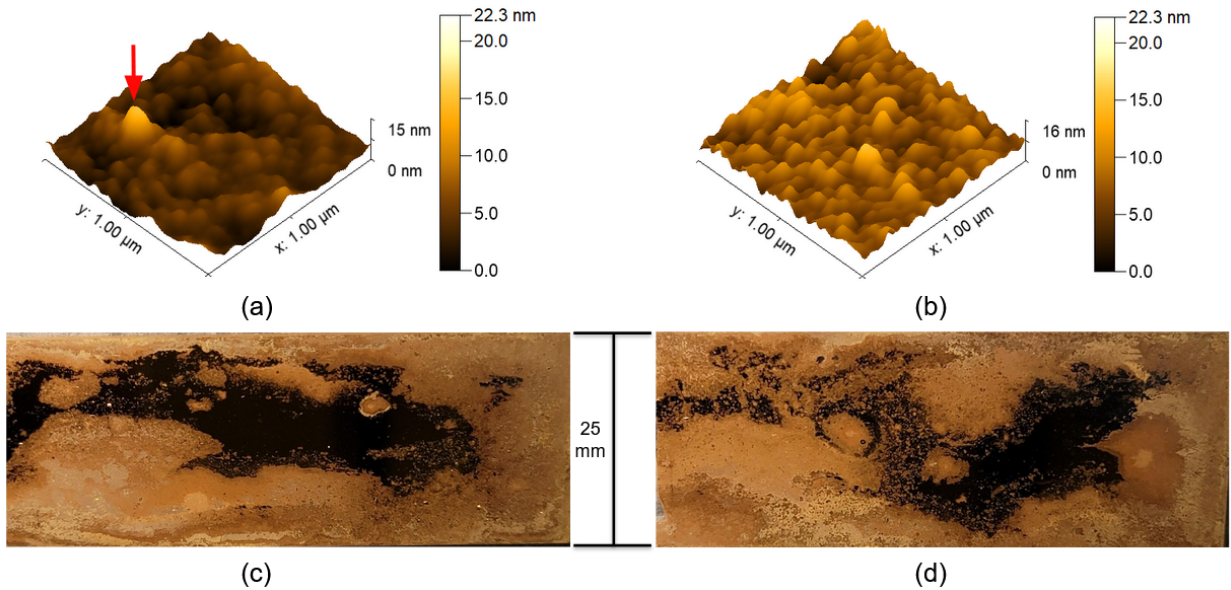

(a)  (b)

(c)  (d)

***Figure 4:*** *(a) Surface morphology of $H_2O(10)$ with RMS roughness of 1.225 nm. (b) Surface morphology of PO(10) with RMS roughness of 1.383 nm. (c) and (d) Photos of the entire area of $H_2O(10)$ and PO(10), respectively, where specular areas appear in darker contrast (i.e., black areas) and corroded areas appear lighter in contrast (i.e., yellow to light brown areas). Both $H_2O(10)$ and PO(10) show a significant smoothing compared to $H_2O(0)$ and PO(0) shown in Figure 3 $H_2O(10)$ underwent the most smoothing with all of the excess material appearing to have been pulled into distinctively tall peaks that appear to be dots on the surface.*

Figure 4(a) and (b) show surface morphology collected by AFM on $H_2O(10)$ and PO(10), respectively. Figure 4(c) and (d) show optical images of the whole samples of $H_2O(10)$ and PO(10), respectively, revealing that significant corrosion took place during the 10 HTHH test cycles (Note: the images shown in panels (a) and (b) were collected, by AFM, from specular areas where no visible corrosion was observed.). Specular areas appear to be black while corroded areas appear to be a light yellow to orange in Figure 4 (c) and (d). RMS roughness reduced from 1.374 of $H_2O(0)$ to 1.225 nm of $H_2O(10)$ and from 1.491 of PO(0) to 1.383 nm of PO(10), indicating that the surface of both samples smoothed by $0.1 \sim 0.15$ nm. PO(10) retained rather similar characteristics seen in Figure 3(b) with the average peak height relative to the average valley depth decreasing by around 17% from PO(0) to PO(10). Conversely, on $H_2O(10)$, a small number of prime peaks – one of such prime peaks is denoted by a red arrow in Figure 4(a) – formed seeming to have height in the range of 1.8-15 nm while heights of the surrounding satellite peaks have been significantly reduced in comparison to peaks in Figure 3(a). Considering that the total amount of material should be conserved (i.e., no material was explicitly added or removed from the samples during the HTHH testing.) on the nominal surface of $H_2O(10)$ and that the satellite peaks surrounding the prime peak have lost not only height but also their definition, it would seem that, on $H_2O(10)$, nearby material was pulled into the prime peak – in other words, material near a prime peak tends to diffuse preferentially into the prime peak – consequently compromising the $AlO_x$ protective coating.

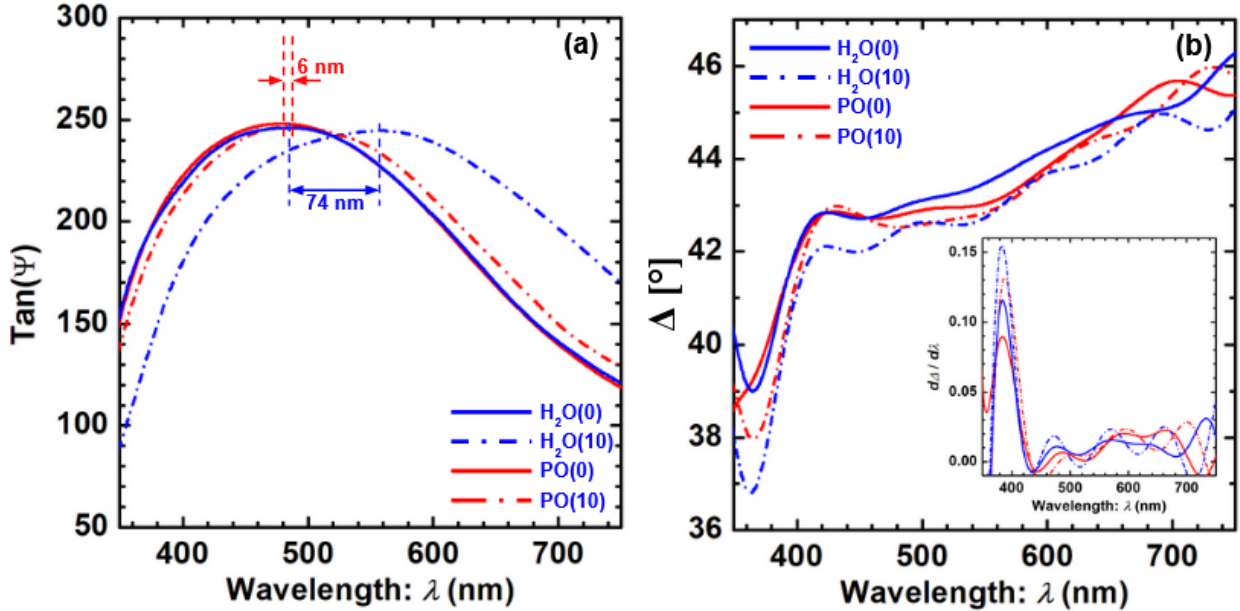

***Figure 5:*** *Two quantities – Tan(Ψ) and Δ – obtained by ellipsometry performed within specular areas of the samples. The ratio of p polarized light to s polarized light – Tan(Ψ) – in $H_2O$ samples shows a significant 74 nm redshift after 10 HTHH cycles while PO samples had a comparatively minimal 6 nm redshift. The change in phase Δ for both was minimal, however the phase change for $H_2O$ was greater and more constant compared to PO where the phase change on average follows the as-prepared sample.*

Figure 5(a) and (b) show *Tan(Ψ)* and phase *Δ*, respectively, obtained by the spectroscopic ellipsometry performed within specular areas where no visible corrosion was observed. Figure 5(a) reveals that $H_2O$(10) underwent significant changes in terms of its surface structure, resulting in a 74 nm shift, while PO(10) had a minimal shift of 6 nm, roughly only 8% that of $H_2O$(10). Figure 5(b) shows minimal changes in *Δ* for both samples: a change at 550 nm by 0.8 degrees for $H_2O$(10) and by XX degrees for PO(10), indicating that the roughening of the surface has little impact on the phase information of the light. These changes in *Tan(Ψ)* and *Δ* are attributed to surface evolution, as seen in Figure 3 and Figure 4 because local variations in film thickness and roughness, as seen in Figure 3 and Figure 4, are expected to significantly impact both *Tan(Ψ)* and *Δ* [11]. The significant implication drawn from Figures 3, 4, and 5 is that specular areas (i.e., areas that do not show obvious macroscopic signs of corrosion–a catastrophic failure) seem to have experienced significant microscopic transformations before reaching a catastrophic failure.

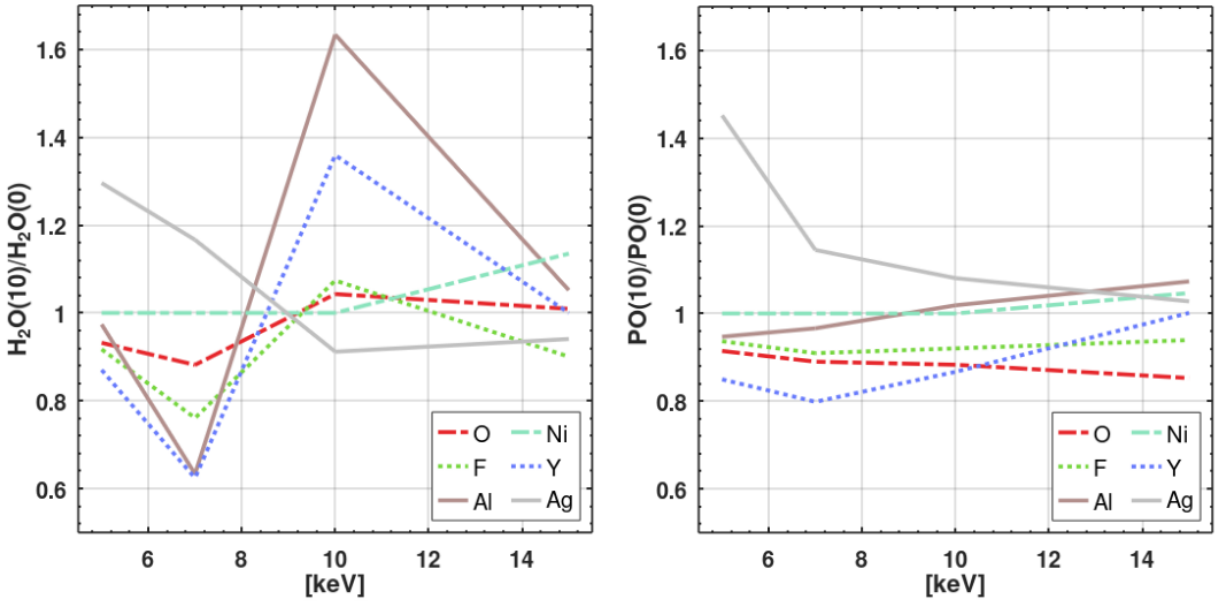

***Figure 6:*** *EDS 10 cycle to 0 cycle ratio at varying electron accelerations (5, 7, 10, and 15 keV). $H_2O$ is very unstable, while PO remains fairly consistent regardless of the depth of the scan.*

The EDS analysis was carried out as follows to obtain information on possible structural rearrangement that took place within the structure below the $AlO_x$ protective coating. First, survey spectra were collected from areas, on $H_2O(0)$, $H_2O(10)$, $PO(0)$, and $PO(10)$, that visibly appeared to be specular (i.e., areas with no visible sign of corrosion) – panels (c) and (d) in Figures 3 and 4 – at various electron beam acceleration voltages. Since the interaction volume in which EDS signals (i.e., characteristic x-rays) are generated depending on the acceleration voltage, EDS signals collected at various acceleration voltages allow us to probe structural changes that occur at various spatial extents. Subsequently, peak intensity $I_j$ of a chemical element $j$ expected to be present in the samples was obtained from the survey spectra of $H_2O(0)$ and $H_2O(10)$ collected at a specific acceleration voltage $V_A$, providing $I_j$ of $H_2O(0)$ and $I_j$ of $H_2O(10)$ for $V_A$. Then, the ratio – $I_j$ of $H_2O(10)$ divided by $I_j$ of $H_2O(0)$ at $V_A$ – was obtained and plotted as a function of $V_A$ in Figure 6(a). The same procedure was used for $PO(0)$ and $PO(10)$ for Figure 6(b). The penetration depths of electron beam generated at various $V_A$ were calculated using the Kanaya-Okayama expression ($R_{KO}$):

$$R_{KO} = 27.6 \, A \, E_o^{1.67}/(\rho \, Z^{0.889}) \tag{1}$$

where $A$ is the average atomic weight of a layer in grams per mol, $E_o$ is the applied electron beam energy in kiloelectron volts, $\rho$ is the density of a layer in grams per cubic-centimeter, and $Z$ is the average atomic number of a layer [14]. The Kanaya-Okayama expression only reveals the maximum penetration depth. Thus, for multiple thin film structures for which the maximum penetration depth exceeds their total thickness, as in our case, the beam intensity needs to be reduced at the interface of the next layer relative to its reduction within the previous layer. As a result, the range of $V_A$ – 5 , 7 , 10 , and 15 keV – used in the experiment, was found to correspond to a range of penetration depth of 250, 483, 1063, and 2341 nm, respectively. Since $H_2O(0)$ and $PO(0)$ are expected to be identical, Figure 6 qualitatively compares structural variations of $H_2O(10)$ and $PO(10)$, directly revealing changes in structural stability of the $H_2O$ and PO samples over the HTHH testing. In other words, even in areas where apparent corrosion has yet to take place, local concentrations of constituent elements change correspondingly and are different from those in the as-prepared samples. For example, in Figure 6 for both $H_2O$ and PO samples, Ag appears to have migrated closer to the samples' surface between the as-prepared samples compared to those after HTHH testing.

In Figure 6, a ratio larger than 1 indicates an increase, in comparison to those in $H_2O(0)$ and $PO(0)$, in the presence of a specific element for a given penetration depth (i.e., a specific interaction volume). A comparison between Figure 6(a) and (b) suggests that, within the subsurface probed by the lowest $V_A$ of 5 keV, both PO and $H_2O$ samples show a general trend – a decrease in the ratio for all elements except Ag. As $V_A$ increases (i.e., as deeper portions are probed and as the interaction volume increases), the ratios for all the elements behave in ways substantially different in the two samples. As seen in Figure 6(a), the $H_2O(10)$ shows a significant drop for all elements except Ag at 7 keV. As $V_A$ further increases to 10 keV, all the elements except Ag increase while Ag drops below 1. By 15 keV the ratios of all the elements appear to be comparable to what they are at 5 keV except for a significant increase in Si, which in turn can be simply explained by the electron beam reaching the BK7 substrates as $V_A$ increased towards 15 keV. In contrast, PO(10) in Figure 6(b) shows an elemental change that remains fairly stable once $V_A$ = 10 keV is reached. There are a few notable features however, Al appears to be reduced at the surface, and increases monotonically as $V_A$ increases. Additionally, a large increase in Ag at $V_A$ = 5 eV is seen, indicating that Ag in the Ag layer and Al in the AlO$_x$ protective coating were able to freely trade positions during the HTHH environment testing. Interestingly, the amount of O and F in the sample appears to have decreased across the board, indicating that these elements were possibly given off as gasses during the HTHH testing. Similarly, Ag decreases as $V_A$ increases, yet Ag remains above 1, seemingly indicating a net increase in Ag within the material, likely in part from the lateral increase in $R_{KO}$ as well as a result of minor variation from the deposition process.

The migration of elements implied by Figure 6(a) and (b), indicates that there is far less of a clear division between layers after the HTHH testing. This comes as no surprise as the samples were previously exposed to at most 60°C during the thermal ALD process, and it seems quite likely that the 80°C environmental testing conditions were enough to allow for noticeable internal reconstruction thermally driven and possibly assisted by higher humidity. This reconstruction may have worked in the favor of the PO sample as in comparison with the $H_2O$ sample – and with the exception of an increase of Ag towards the surface of both samples – the PO sample remained fairly consistent with minimal apparent changes to the remaining structure, resulting in the PO sample outperforming the $H_2O$ sample as an oxygen precursor in the preparation of an $AlO_x$ protective coating.

## 4. Conclusion

Ag-based telescope mirrors require optically transparent protective coatings to function for extended periods of time without affecting the mirror's functionality. While $AlO_x$ is one of the more viable options for a protective coating, how it is deposited has a significant impact on the ultimate lifespan of the mirror. The conventional method of using $H_2O$ as an oxygen precursor works well for many applications; however, as our study clearly suggests, the use of PO would offer many benefits that water is unable to do. For our specific objective – durable Ag-based mirrors – our study may lead to durable coatings that maintain the original optical performance for a much longer operating lifetime. When comparing between specular areas, PO had a minimal 6 nm shifting in s to p polarization compared to the 74 nm shift $H_2O$ suffers from, indicating that, over long term use and despite appearing uncorroded, conventionally protected Ag mirrors will have significant variations in observational outcomes even when looking at the same astronomical object. This conclusion is also supported by the EDS results that revealed the $H_2O$ sample had rather erratic structural changes during the HTHH test cycles, while the PO sample maintained structural integrity. This may result in the surface of the PO sample gradually degrading, as indicated by the relative consistency between the as-prepared and HTHH tested PO samples' AFM data, while the surface of the $H_2O$ sample was compromised by the formation of tall peaks that absorbed surrounding material, fundamentally weakening the protective coating. Therefore, the PO sample has conclusively proven to have significantly more durability leading to a 17% improvement in reflectance spectra in comparison to that of the $H_2O$ sample.

**Acknowledgement**

This study was partially supported by Dr. John Hennessy at Jet Propulsion Laboratory, California Institute of Technology (Pasadena, California).

**References**

**References**

[1] D. M. Fryauf, A. C. Phillips, N. P. Kobayashi, Annealing of high performance silverbased mirrors, in: N. P. Kobayashi, A. A. Talin, A. V. Davydov (Eds.), Low-Dimensional Materials and Devices 2021, Vol. 11800, International Society for Optics and Photonics, SPIE, 2021, p. 1180008. doi:10.1117/12.2595147. URL https://doi.org/10.1117/12.2595147

[2] D. M. Fryauf, A. C. Phillips, N. P. Kobayashi, Critical processing temperature for highperformance protected silver mirrors, Journal of Astronomical Telescopes, Instruments, and Systems 7 (3) (2021) 034002. doi:10.1117/1.JATIS.7.3.034002. URL https://doi.org/10.1117/1.JATIS.7.3.034002

[3] F. Anjum, D. M. Fryauf, R. Ahmad, A. C. Phillips, N. P. Kobayashi, Improving silver mirrors with aluminum oxynitride protection layers: variation in refractive index with controlled oxygen content by radiofrequency magnetron sputtering, Journal of Astronomical Telescopes, Instruments, and Systems 4 (4) (2018) 044004. doi:10.1117/1.JATIS.4.4.044004. URL https://doi.org/10.1117/1.JATIS.4.4.044004

[4] D. M. Fryauf, J. J. D. Leon, A. C. Phillips, N. P. Kobayashi, Effect of intermediate layers on atomic layer deposition-aluminum oxide protected silver mirrors, Journal of Astronomical Telescopes, Instruments, and Systems 3 (3) (2017) 034001. doi:10.1117/1.JATIS.3.3.034001. URL https://doi.org/10.1117/1.JATIS.3.3.034001

[5] D. M. Fryauf, A. C. Phillips, N. P. Kobayashi, Corrosion protection of silver-based telescope mirrors using evaporated anti-oxidation overlayers and aluminum oxide films by atomic layer deposition, in: N. P. Kobayashi, A. A. Talin, M. S. Islam, A. V. Davydov (Eds.), Low-Dimensional Materials and Devices 2016, Vol. 9924, International Society for Optics and Photonics, SPIE, 2016, p. 99240S. doi:10.1117/12.2238749. URL https://doi.org/10.1117/12.2238749

[6] D. M. Fryauf, A. C. Phillips, N. P. Kobayashi, Corrosion barriers for silver-based telescope mirrors: comparative study of plasma-enhanced atomic layer deposition and reactive evaporation of aluminum oxide, Journal of Astronomical Telescopes, Instruments, and Systems 1 (4) (2015) 044002. doi:10.1117/1.JATIS.1.4.044002. URL https://doi.org/10.1117/1.JATIS.1.4.044002

[7] D. M. Fryauf, A. C. Phillips, N. P. Kobayashi, Moisture barrier and chemical corrosion protection of silver-based telescope mirrors using aluminum oxide films by plasmaenhanced atomic layer deposition, in: N. P. Kobayashi, A. A. Talin, A. V. Davydov, M. S. Islam (Eds.), Nanoepitaxy: Materials and Devices V, Vol. 8820, International Society for Optics and Photonics, SPIE, 2013, p. 88200Y. doi:10.1117/12.2023826. URL https://doi.org/10.1117/12.2023826

[8] R. W. Johnson, A. Hultqvist, S. F. Bent, A brief review of atomic layer deposition: from fundamentals to applications, Materials Today 17 (5) (2014) 236–246. doi:https:

//doi.org/10.1016/j.mattod.2014.04.026. URL

https://www.sciencedirect.com/science/article/pii/S1369702114001436

[9] High-concentration/high-purity ozone gas generator: Pure Ozone Generator (9 2022). URL https://www.meidensha.com/catalog/MB64-3059.pdf

[10] OG-5000 Series High Purity Ozone Generator (11 2017). URL www.teledyne-api.com/prod/Downloads/SAL000099C%20-%20OG-5000.pdf

[11] B. Hajduk, H. Bednarski, B. Trzebicka, Temperature-dependent spectroscopic ellipsometry of thin polymer films, The Journal of Physical Chemistry B 124 (16) (2020) 3229–3251, pMID: 32275433. arXiv:https://doi.org/10.1021/acs.jpcb.9b11863, doi:10.1021/acs.jpcb.9b11863. URL https://doi.org/10.1021/acs.jpcb.9b11863

[12] D. Neˇcas, P. Klapetek, Gwyddion: an open-source software for SPM data analysis, Central European Journal of Physics 10 (2012) 181–188. doi:10.2478/s11534-011-0096-2.

[13] J. W. Eaton, many others, Gnu octave, version 6.1.0, https://www.gnu.org/software/octave/ (2020).

[14] J. I. Goldstein, D. E. Newbury, P. Echlin, D. C. Joy, C. Fiori, E. Lifshin, ElectronBeam-Specimen Interactions, Springer US, Boston, MA, 1981, pp. 53–122. doi:10.1007/978-1-4613-3273-2_3. URL https://doi.org/10.1007/978-1-4613-3273-2_3